\documentclass[preprint,12pt]{elsarticle}

\usepackage{xcolor}
\usepackage{epsfig}
\usepackage{amsmath,amssymb,bm}
\usepackage{enumerate}
\usepackage{graphicx}
\usepackage{color}
\usepackage{braket}
\usepackage{hyperref}
\usepackage{lineno}
\usepackage{booktabs}
\usepackage{xspace}
\usepackage{framed}

\def\be{\begin{equation}}
\def\ee{\end{equation}}

\makeatletter
\def\ps@pprintTitle{%
   \let\@oddhead\@empty
   \let\@evenhead\@empty
   \def\@oddfoot{\centerline{\thepage}}%
   \let\@evenfoot\@oddfoot}
\makeatother

\begin{document}

\begin{frontmatter}
\title{Searches for extra-dimensional excitations in light-by-light scattering}

\author{Malak Ait Tamlihat$^{\dagger,a,\star}$, Ghizlane Ez-Zobayr$^{b,\star}$, Laurent Schoeffel$^{c,\dagger\dagger}$, Yahya~Tayalati$^{a,b,\mathsection}$}
\address{$\dagger$ Authors in alphabetical order}
\vspace{5mm}
\address{$a$ Mohammed V University in Rabat, Faculty of Sciences, 4 av. Ibn Battouta, B.P. 1014, R.P. 10000 Rabat, Morocco\\
$b$ School of Physics Applied and Engineering, Mohammed VI Polytechnic University, Lot 660, 43150 Hay Moulay Rachid Ben Guerir, Morocco \\
$c$ Irfu, CEA, Université Paris-Saclay, 91191 Gif-sur-Yvette, France
}
\vspace{20mm}
\address{$^\star$malak.ait.tamlihat@cern.ch, $^\star$ghizlane.ez-zobayr@cern.ch, 
$^{\dagger\dagger}$laurent.olivier.schoeffel@cern.ch, $^\mathsection$Yahya.Tayalati@cern.ch.}

\begin{abstract}
\noindent
We present a comprehensive phenomenological analysis of the Radion in the Randall-Sundrum model, focusing on its production via light-by-light scattering in ultra-peripheral proton-proton collisions at the LHC. We provide a consistent derivation of the effective couplings to Standard Model fields, clarifying the normalization of the trace anomaly-induced coupling to photons and the role of kinetic mixing with the Higgs boson. We demonstrate that while the pure gravitational coupling is loop-suppressed relative to Axion-Like Particles (ALPs), making the unmixed Radion elusive, the non-minimal mixing with the Higgs sector can induce constructive interference that enhances the signal by orders of magnitude. Using forward proton tagging to select exclusive high-mass events, we reinterpret recent experimental limits on ALPs to derive the first exclusion contours for the Radion in the $(\Lambda_r, \xi)$ plane, showing that mixing scenarios are beginning to be constrained by current LHC data.
\end{abstract}

\end{frontmatter}

\section{Introduction}

\noindent
In the Standard Model (SM) of particle physics, the hierarchy problem refers to the huge gap between two energy scales that emerges in the theory, namely the Planck mass scale $M_{Pl} \simeq 10^{19}$ GeV and the electroweak scale characterized by the vacuum expectation value for the Higgs field $v \approx 246 \text{ GeV}$. In addition, the SM suggests that the Higgs boson mass should be driven up to the Planck scale by quantum corrections, requiring an impossibly precise fine-tuning to keep it at its observed value of $m_H \approx 125 \text{ GeV}$. More precisely, we can write
\be
\label{eq1}
-\frac{m_{H}^2}{2} = m_0^2 + \delta m_H^2,
\ee
where $m_0^2$ is the bare mass squared and $\delta m_H^2$ represents the quadratically divergent quantum corrections
$$
\delta m_H^2 \simeq -C M_{Pl}^2 \simeq - (10^{18} \text{ GeV})^2.
$$
This necessitates that $m_0^2$ be an immensely large positive number, fine-tuned to cancel the quantum corrections to an arbitrary precision to lead to the physical mass squared. This is the fine-tuning crisis immediately linked to the hierarchy problem that appears in the SM.

Extra Dimensional (ED) models, and in particular the Randall-Sundrum (RS) model \cite{Randall:1999ee,Csaki:2007ns,Giudice} where the Radion field arises, offer elegant solutions to solve the hierarchy problem and the related fine-tuning issue. The RS model addresses the problem by introducing a warped fifth dimension ($y$) that naturally generates the hierarchy. Indeed, the RS metric reads
\be
\label{rs}
ds^2 = {e^{-2ky}} \eta_{\mu\nu}dx^\mu dx^\nu + dy^2,
\ee
where $\eta_{\mu\nu} = \text{diag}(1, -1, -1, -1)$ is the 4D Minkowski metric tensor and $k>0$ is the curvature scale, with $k \sim M_{Pl}$. This geometry generates two 4D branes:
\begin{itemize}
\item[(1)] The Planck-brane ($y=0$) where gravity is fundamentally strong. The fundamental energy scale of physics on this brane is the Planck mass.
\item[(2)] The TeV-brane ($y=y_c=\pi r_c$), where $r_c \sim L_{Pl}=10^{-35}$ m (a few Planck lengths) is the compactification radius. This corresponds to the 4D space where the SM resides, our world.
\end{itemize}

The warp factor $e^{-ky_c}$ that appears on the TeV-brane (see Eq. (\ref{rs})) provides the solution to the hierarchy problem. Indeed, it can be shown that the fundamental mass scale $m_0 \sim M_{Pl}$ on the Planck-brane is perceived on the TeV-brane as an exponentially smaller warped mass
$$
m_{\text{phys}} = m_0 \times e^{-ky_c},
$$
which can be of order $100$ GeV by choosing $ky_c \simeq 39$ (see Appendix). Achieving this value is natural, as
$$
k y_c \sim \frac{1}{M_{Pl}} \pi L_{Pl}.
$$

We note that while there are obviously still quantum loop corrections in the RS model, the cut-off scale on the TeV-brane is not $M_{Pl}$ but $e^{-ky_c} M_{Pl}$. This means that these corrections are naturally TeV-sized rather than Planck-sized. Therefore, returning to Eq. (\ref{eq1}), there is no need for fine-tuning.

In this context, the Radion field arises when the size of the ED, specifically the distance between the two branes, is not fixed but is allowed to fluctuate. In this Letter, we perform a comprehensive phenomenological analysis of the Radion at the LHC, clarifying the theoretical normalization of its couplings which is of crucial importance for this work as demonstrated in Section 6. In Section 2, we review the geometric origin of the Radion field and establish the consistency between different gauge choices. Section 3 provides a detailed derivation of the effective couplings to Standard Model fields, with a specific focus on the trace anomaly-induced coupling to photons and the kinetic mixing with the Higgs boson. We demonstrate that the interplay between the conformal anomaly and the Higgs-mixing contributions is critical for the phenomenology. In Section 4 and 5, we simulate the production of the Radion via light-by-light scattering in proton-proton collisions, utilizing forward proton tagging to suppress backgrounds. Finally, in Section 6, we reinterpret existing experimental limits on Axion-Like Particles (ALPs) to derive exclusion contours for the Radion in the mass-coupling plane, highlighting the significant suppression of the pure gravitational signal relative to the ALP case.

\section{Expression of the Radion as a scalar field in the TeV-brane}

The geometry of the Randall-Sundrum model is defined on a 5D spacetime where the extra spatial dimension is compactified on an orbifold $S^1/\mathbb{Z}_2$. Geometrically, this corresponds to a circle of radius $r_c$ (with coordinate $y \in [-\pi r_c, \pi r_c]$) subjected to the orbifold identification $y \leftrightarrow -y$. The resulting physical space is a line segment (the fundamental domain) $y \in [0, \pi r_c]$. The boundaries of this compactified dimension are the fixed points of the $\mathbb{Z}_2$ symmetry:
\begin{itemize}
    \item The Planck-brane located at $y=0$.
    \item The TeV-brane located at $y = y_c \equiv \pi r_c$.
\end{itemize}

The Radion field arises because the geometry of this extra dimension is dynamical. Specifically, the compactification radius $r_c$ (and thus the inter-brane distance $y_c$) is not fixed but is allowed to fluctuate. The idea is that this distance ($y_c \rightarrow y_c(x)$) is promoted to a dynamical 4D field, the Radion, which represents the fluctuations of the orbifold size. As in 4D General Relativity, a key concept in the RS 5D model is 5D diffeomorphism invariance. In practice, this means that there is a gauge freedom allowing us to choose how to represent this fluctuation of the compactified dimension in the 5D metric.

\subsection{The Unitary Gauge}

The most simple gauge is what is called the {Unitary Gauge}. In this picture, the coordinate length of the orbifold interval is kept rigid ($y$ runs from $0$ to a fixed $y_c$), while the fluctuation of the geometry is placed entirely into the components of the metric tensor. In the full 5D formalism, solving the linearized Einstein equations in the bulk requires the metric to take the specific form
\be
\label{ugrs}
ds^2 = e^{-2k y} \left[ (1 - 2F(x)) \eta_{\mu\nu} \right] dx^\mu dx^\nu + (1 + 4F(x)) dy^2.
\ee
Here, $F(x)$ is the dimensionless Radion perturbation. In the following, we treat the fluctuation of the ED as a perturbation, implying $|F(x)| \ll 1$. Note the interplay between the components: the term $(1+4F(x))$ in $g_{55}$ is required to keep the bulk geometry Ricci-flat ($R_{MN}=0$) between the branes.
We rewrite Eq. (\ref{ugrs}) as
$$
ds^2 = e^{-2k y} \underbrace{\left[ \Omega^2(x) \eta_{\mu\nu} \right]}_{G_{\mu\nu}(x)} dx^\mu dx^\nu + G_{yy}(x) dy^2,
$$
with
$$
    \Omega^2(x) = 1+ \frac{2\phi(x)}{\Lambda_\phi}.
$$
Here, $\phi(x)$ is the canonical scalar field, called the Radion field, function of the 4D spacetime coordinates $x^\mu$, and $\Lambda_\phi$ is a fundamental coupling scale for the Radion in the 5D RS theory, thus of the order of the Planck mass.

{\it Remark on the sign convention:} It is important to clarify the relationship between the bulk perturbation $F(x)$ and the field $\phi(x)$. In the linearized bulk metric, the 4D term appears as $(1 - 2F(x))$ (see Eq. (\ref{ugrs})), carrying a minus sign. This reflects the physical nature of the warped geometry. A positive fluctuation in the size of the extra dimension ($g_{55}$ increases) effectively increases the distance $y$, which in turn \textit{decreases} the warp factor $e^{-2ky}$. In our definition of $\Omega^2(x)$, we translate this into a positive sign for the scalar field $(1+2\phi/\Lambda_\phi)$. This implies that the canonical field $\phi(x)$ is defined to track the conformal factor on the brane, rather than the geometric distance directly. As we show below, this convention ensures consistency when transforming between different gauges.

Then, in order to find the 4D metric $g_{\mu\nu}(x)$ of the SM (on the TeV-brane), we evaluate the 4D part of the 5D metric ($G_{\mu\nu}$) at the fixed orbifold boundary $y=y_c$. Since we are restricted to the brane, the $dy^2$ term vanishes. We get
$$
g_{\mu\nu}(x) = G_{\mu\nu}(x, y=y_c) = e^{-2ky_c} \left[1+\left(\frac{2\phi(x)}{\Lambda_\phi}\right) \eta_{\mu\nu} \right].
$$
With the notation $g_{\mu\nu}^{(0)} = e^{-2ky_c} \eta_{\mu\nu}$, we can rewrite
\be
\label{Radion1}
g_{\mu\nu}(x) =  \left[1+\left(\frac{2\phi(x)}{\Lambda_\phi}\right)  \right]g_{\mu\nu}^{(0)}.
\ee
In this gauge, the position of the TeV-brane in the compact dimension is static, but the ruler (the metric) on the brane is dynamically stretching and shrinking, as dictated by the scale $\Omega(x) =  1+{\frac{\phi(x)}{\Lambda_\phi}}$.

\subsection{The Brane Bending Gauge}

The other standard gauge choice (often called the {Brane Bending Gauge} or Gaussian Normal coordinates) is to consider directly the fluctuation of the TeV-brane position. The fluctuation is then pushed entirely into the dynamical position of the boundaries of the orbifold. The interest is that this is more physically intuitive from the 5D perspective. The 5D metric in the bulk is like the original RS vacuum metric
$$
ds^2 =  e^{-2k y} \eta_{\mu\nu} dx^\mu dx^\nu + dy^2,
$$
but where the TeV-brane is no longer kept at the fixed compactification radius $y_c = \pi r_c$. Instead, the boundary is displaced:
$$
y(x) = y_c + \pi(x),
$$
where the 4D scalar field $\pi(x)$ is the Radion field in this gauge.
In order to derive the 4D effective metric on the TeV-brane, we must evaluate the 5D metric at the brane dynamical position $y(x)$. Substituting $dy = \partial_\mu \pi dx^\mu$ into the metric generates the kinetic term for the Radion. However, for the conformal scaling of the matter fields, we look at the induced metric factor from the warp factor
$$
g_{\mu\nu}(x) =  e^{-2k y(x)} \eta_{\mu\nu}= e^{-2ky_c} e^{-2k\pi(x)} \eta_{\mu\nu}.
$$
Again, using $g_{\mu\nu}^{(0)} = e^{-2ky_c} \eta_{\mu\nu}$, we obtain
\be
\label{Radion2}
g_{\mu\nu}(x) = e^{-2k\pi(x)} g_{\mu\nu}^{(0)}
\ee
Here, it is the TeV-brane which is physically moving up and down in the warped fifth dimension, as dictated by the 4D Radion field $\pi(x)$.

\subsection{The Modulus and Brane Bending Gauges}

In the literature (e.g. Ref. \cite{Csaki:2007ns}), the Radion is often introduced by promoting the compactification radius $r_c$ itself to a dynamical field $r(x)$. Using the angular coordinate $\varphi$ of the orbifold circle (where $y = r_c \varphi$), the metric takes the form:
\be
\label{modulus_metric}
ds^2 = e^{-2k r(x) \varphi} \eta_{\mu\nu} dx^\mu dx^\nu + r(x)^2 d\varphi^2.
\ee
In this gauge, the boundaries are fixed at the orbifold points $\varphi=0$ and $\varphi=\pi$, but the proper length of the extra dimension fluctuates locally as $L(x) = \pi r(x)$. This is known as the {Modulus Parametrization}.

This form is equivalent to the {Brane Bending Gauge} (or Gaussian Normal coordinates) used in our Section 2.1. To see this, consider the coordinate transformation from the angular frame $(\varphi)$ to the proper distance frame $(y)$:
$$
y(x, \varphi) = r(x) \varphi.
$$
If we assume the fluctuation is small, the differential transforms as $dy \approx r(x) d\varphi$ (neglecting the $\partial_\mu r$ terms which are higher order in the derivative expansion). Substituting this into Eq. (\ref{modulus_metric}), the $g_{55}$ component becomes unitary:
$$
r(x)^2 d\varphi^2 \rightarrow dy^2.
$$
However, the boundary condition at $\varphi = \pi$, which was fixed in the modulus gauge, now becomes a fluctuating boundary in the $y$-coordinate:
$$
y_{brane}(x) = r(x) \pi \equiv y_c + \pi(x).
$$
Thus, we recover the {Brane Bending} picture where the bulk metric is static, but the TeV-brane fluctuates:
\be
\label{brane_bending}
ds^2 = e^{-2ky} \eta_{\mu\nu} dx^\mu dx^\nu + dy^2, \quad \text{with brane at } y = y_c + \pi(x).
\ee
The physical induced metric on the brane is identical in both cases. In the modulus gauge (Eq. \ref{modulus_metric}), evaluating at the fixed $\varphi=\pi$ gives:
$$
g_{\mu\nu}^{ind} = e^{-2k r(x) \pi} \eta_{\mu\nu} = e^{-2k(y_c + \pi(x))} \eta_{\mu\nu}.
$$
In the brane bending gauge (Eq. \ref{brane_bending}), evaluating at the moving $y = y_c + \pi(x)$ gives the exact same result:
$$
g_{\mu\nu}^{ind} = e^{-2k(y_c + \pi(x))} \eta_{\mu\nu}.
$$
This confirms that the scalar field $\pi(x)$ (used here) and the dynamical radius $r(x)$ (used in the paper) describe the same physical degree of freedom.

\subsection{Proof of equivalence between gauges}

It is instructive to demonstrate explicitly that these two gauges describe the same physical reality through a coordinate transformation. We start from the {Brane Bending gauge}, where the brane is displaced
$$
ds^2 = e^{-2ky} \eta_{\mu\nu} dx^\mu dx^\nu + dy^2, \quad \text{with brane at } y = y_c + \pi(x).
$$
To move to the {Unitary gauge}, we seek a new coordinate system $(\tilde{x}, \tilde{y})$ where the brane position is fixed at the orbifold boundary $\tilde{y} = y_c$. We perform a coordinate transformation of the ED:
\be
\label{coord_trans}
y = \tilde{y} + \pi(x).
\ee
In the original coordinates, the brane is at $y(x) = y_c + \pi(x)$. Substituting this into Eq.~(\ref{coord_trans}), we get:
$$
y_c + \pi(x) = \tilde{y} + \pi(x) \quad \Longrightarrow \quad \tilde{y} = y_c.
$$
This corresponds precisely to the straight brane definition of the Unitary gauge.
Now, we must determine how the metric tensor transforms. Substituting Eq.
(\ref{coord_trans}) into the warp factor $e^{-2ky}$, we obtain
$$
e^{-2ky} = e^{-2k(\tilde{y} + \pi(x))} = e^{-2k\tilde{y}} e^{-2k\pi(x)}.
$$
Assuming the fluctuation $\pi(x)$ is small, we can Taylor expand the exponential term containing the field
$$
e^{-2k\pi(x)} \approx 1 - 2k\pi(x).
$$
Thus, the 4D part of the metric transforms as
$$
ds^2_{(4D)} \approx e^{-2k\tilde{y}} \left[ (1 - 2k\pi(x)) \eta_{\mu\nu} \right] dx^\mu dx^\nu.
$$
We now compare this result with the definition of the metric in the Unitary gauge given in Eq.
(\ref{Radion1})
$$
ds^2_{(4D)} = e^{-2k\tilde{y}} \left[ \left( 1 + \frac{2\phi(x)}{\Lambda_\phi} \right) \eta_{\mu\nu} \right] dx^\mu dx^\nu.
$$
By identifying the coefficients of $\eta_{\mu\nu}$ in the two expressions, we find the direct relationship between the brane displacement $\pi(x)$ and the bulk scalar field $\phi(x)$:
$$
-2k\pi(x) = \frac{2\phi(x)}{\Lambda_\phi} \quad \Longrightarrow \quad \phi(x) = - (k \Lambda_\phi) \pi(x).
$$
This proves that the scalar field $\phi(x)$ in the Unitary gauge effectively captures the physical displacement of the brane from the Brane Bending gauge. The {\it wiggle} of the brane has been absorbed by the metric tensor, generating the Radion field $\phi(x)$.

\subsection{$\phi(x)$ is a scalar field}
Let us now give the main elements to prove that $\phi(x)$ is a scalar field, with a Lagrangian density of the form
\be
\label{Radionmain}
\mathcal{L}_{\text{Radion}} = \frac{1}{2} \partial_\mu \phi(x) \partial^\mu \phi(x) - V(\phi).
\ee
In this expression, the potential $V(\phi)$ must have a minimum so that the size of fluctuation is constrained and the second derivative of $V(.)$ trivially gives the mass to the Radion
$$
m_\phi^2 = \frac{d^2V}{d\phi^2}|_{\phi=0}.
$$

{\it Proof:}
we start from the 5D RS action
$$
S_5 = 2M_5^3 \int d^5x \sqrt{-G} \ R_5,
$$  
where, $M_5$ is the 5D fundamental  scale of the theory, thus of the order of the Planck mass, $G$ is the determinant of the 5D RS metric $G_{MN}$ and $R_5$ is the 5D Ricci scalar curvature. This scalar curvature is computed from the above RS metric.

In this proof, we use the Unitary gauge defined above. To extract the kinetic term, we proceed with the simplified ansatz where the dynamical degree of freedom is projected onto the 4D scale factor $\Omega(x)$.
$$
ds^2 = e^{-2ky} \left[ \Omega^2(x) \eta_{\mu\nu} \right] dx^\mu dx^\nu + dy^2.
$$    
The determinant of the metric reads
$$
G = \det(G_{\mu\nu}) \times G_{yy} = (-e^{-8ky}\Omega^8(x)) \times (1) = -e^{-8ky}\Omega^8(x),
$$
which gives
$$ 
\sqrt{-G} =  \sqrt{-(-e^{-8ky} \Omega^8(x))} =  \sqrt{e^{-8ky} \Omega^8(x)} = e^{-4ky} \Omega^4(x).
$$
Then, the 5D Ricci scalar reads
$$
R _5 = e^{2ky} R_4 \left[ \Omega^2(x) \eta_{\mu\nu} \right] + ( \text{terms with only } y \text{-derivatives})
$$
The $y$-derivative terms are part of the static background solution and contain no 4D derivatives, so we can ignore them for the kinetic derivation. We are left with the kinetic part of $R_5$
$$
R _5^{ \text{kin}} = e^{2ky} R_4 \left[ \Omega^2(x) \eta_{\mu\nu} \right],
$$
where the 4D Ricci scalar $R_4$ is a standard calculus
$$
R_4 \left[ \Omega^2(x) \eta_{\mu\nu} \right] = -6  \Omega^{-3} (x) \Box  \Omega(x).
$$
We use the notation $\Box \Omega(x) \equiv \eta^{\mu\nu}\partial_\mu \partial_\nu \Omega(x)$. This leads to
$$
R _5^{ \text{kin}} = e^{2ky}  \left( -6  \Omega^{-3} (x)  \Box  \Omega(x)  \right).
$$
Now we substitute these into the kinetic part of the action and integrate on the fifth dimension
$$
S_5^{\text{kin}} = 2M_5^3 \int d^4x \int_0^{y_c} dy \left( \sqrt{-G} \right) \left( R_5^{\text{kin}} \right),
$$
or
$$
S_5^{\text{kin}} = 2M_5^3 \int d^4x \int_0^{y_c} dy \left( e^{-4ky}\Omega^4(x) \right) \left( e^{2ky} (-6 \Omega^{-3} (x) \Box \Omega(x)) \right).
$$
Combining the terms, we obtain
$$
S _5^{ \text{kin}} = 2M _5^3  \int d^4x  \left( \int_0^{y_c} e^{-2ky} dy \right)  \left( -6  \Omega(x)  \Box  \Omega(x)  \right).
$$
The integral over $y$ is a constant geometric factor:
$$
\int_0^{y_c} e^{-2ky} dy = \left[ \frac{e^{-2ky}}{-2k} \right]_0^{y_c} = \frac{1 - e^{-2ky_c}}{2k} \approx \frac{1}{2k},
$$
where we used the hierarchy assumption $e^{-2ky_c} \ll 1$. At this step, the action becomes:
$$
S _5^{ \text{kin}} = -  \left(  \frac{6 M _5^3}{k}  \right)  \int d^4x ( \Omega(x)  \Box  \Omega(x))
$$
Finally, in order
to recover the standard kinetic term $(\partial \Omega)^2$, we perform an integration by parts on the 4D integral:
$$
\int d^4x \, \Omega(x) \Box \Omega(x) = \underbrace{\int d^4x \, \partial_\mu (\Omega \partial^\mu \Omega)}_{\text{boundary term } \to 0} - \int d^4x \, (\partial_\mu \Omega(x) \partial^\mu \Omega(x)).
$$
Substituting this back into the action flips the sign:
$$
S _5^{ \text{kin}} =  \left(  \frac{6 M _5^3}{k}  \right)  \int d^4x ( \partial_\mu  \Omega(x)  \partial^ \mu  \Omega(x))
$$
 The effective 4D Lagrangian density then reads as a function of 
$\Omega(x)$
$$
\mathcal{L}[\Omega(x)] = \left( \frac{6 M_5^3}{k} \right) (\partial_\mu \Omega(x) \partial^\mu \Omega(x)).
$$
Finally, it is straightforward to express this in terms of $\phi(x)$ in the approximation of small fluctuations 
$$
\Omega(x) =  1 + \phi(x) / \Lambda_\phi.
$$
With $\partial_\mu \Omega(x) \approx \frac{1}{\Lambda_\phi} \partial_\mu \phi$, we finally obtain
\be
\label{kin1}
 \mathcal{L}[ \phi] =  \left[  \frac{6 M _5^3}{k  \Lambda_\phi^2}  \right] ( \partial_\mu  \phi  \partial^ \mu  \phi)
\ee
We have the freedom to define the scale $\Lambda_\phi$ such that the prefactor is $\frac{1}{2}$ by taking 
\be
\label{lambda1}
\Lambda_\phi^2 = \frac{12 M_5^3}{k}.
\ee
Let us notice that we confirm with Eq. (\ref{lambda1}) that $\Lambda_\phi$ is a fundamental scale in the theory that is thus of the order of the Planck mass. Finally, we obtain
$$
\mathcal{L}_{\text{Radion}} =  \mathcal{L}[ \phi] = \frac{1}{2} \partial_\mu \phi(x) \partial^\mu \phi(x) 
$$

\section{Coupling of Radion to SM particles}

The Radion couples to matter particles because, in its essence, it is a component of the gravitational field. Its interaction with matter inherits from the principle of equivalence in General Relativity (GR), extended to higher dimensions. In GR, matter fields couple to the spacetime geometry described by the metric $g_{\mu\nu}$ and the source for gravity is the energy-momentum tensor $T_{\mu\nu}$. Then, the interaction term of the scalar field $\phi(x)$ with matter (on the TeV-brane) is necessarily proportional to the trace of this tensor 
\be
\label{int1}
\mathcal{L}_{int} = \frac{\phi}{\Lambda_r} T^\mu_\mu,
\ee
where $\Lambda_r$ is the Radion scale, but this time perceived on the TeV-brane, thus $\Lambda_r \simeq \Lambda_\phi e^{-ky_c}$ is a TeV scale. The argument is similar to the result of the Appendix  but the proof requires to vary the action with respect to the metric.
{\it Proof:} let us write the metric on the TeV-brane as a perturbation around the background. From Eq.
(\ref{Radion1}) in the previous section, we have:
$$
g_{\mu\nu}(x) = e^{-2ky_c} \eta_{\mu\nu} \left( 1 + 2\frac{\phi(x)}{\Lambda_\phi} \right).
$$
The action for SM fields $\psi_{SM}$  the TeV-brane depends on this induced metric:
$$
S_{SM} = \int d^4x \sqrt{-g} \ \mathcal{L}_{SM}(g_{\mu\nu}, \psi_{SM}).
$$
The interaction Lagrangian is derived by varying the action with respect to the metric fluctuations caused by the Radion. By definition of the energy-momentum tensor $T^{\mu\nu} = \frac{2}{\sqrt{-g}} \frac{\delta S_{SM}}{\delta g_{\mu\nu}}$, the variation of the action is:
$$
\delta S_{int} = \frac{1}{2} \int d^4x \sqrt{-g} \ T^{\mu\nu} \delta g_{\mu\nu}.
$$
Substituting the metric perturbation $\delta g_{\mu\nu} = e^{-2ky_c} \eta_{\mu\nu} (2\frac{\phi}{\Lambda_\phi})$, and noting that the indices of $T^{\mu\nu}$ are raised/lowered with the background metric, we find that the term leads to the trace $T^\mu_\mu$. Defining the warped scale $\Lambda_r \equiv \Lambda_\phi e^{-ky_c}$, we obtain the canonical interaction term:
$$
\mathcal{L}_{int} = \frac{\phi(x)}{\Lambda_r} T_{\mu}^{\mu(SM)},
$$
which concludes the proof.

\subsection{Coupling to Fermions}
The coupling to matter depends on the trace of the energy-momentum tensor. For fermions, represented by their spinors $\psi$, the trace is proportional to the mass term that breaks scale invariance explicitly:
$$
T^\mu_{\mu(\psi)} = m_\psi \bar{\psi}\psi.
$$
Substituting this into our general interaction formula, we find:
$$
\mathcal{L}_{\phi\psi\psi} = \frac{m_\psi}{\Lambda_r} \phi \bar{\psi}\psi.
$$
In particular, the Radion interacts much more strongly with heavier fermions (like the top quark). This coupling is not an arbitrary new force; it arises because the Radion is a scalar component of gravity. Just as gravity couples to mass-energy, the Radion couples to the part of the Lagrangian density in the SM that breaks scale invariance which, for fermions, is precisely their mass.

\subsection{Coupling to Photons and Trace Anomaly}
Now, let us compute the coupling of the Radion to photons. For the electromagnetic (EM) field, represented by the field strength tensor $F_{\mu\nu}$, the classical energy-momentum tensor is given by:
$$
T^{\mu\nu}_{EM} = F^{\mu\alpha}F^{\nu}_{\alpha} - \frac{1}{4}g^{\mu\nu}F_{\alpha\beta}F^{\alpha\beta}.
$$
If we compute the trace of this tensor using classical electrodynamics, we find that it vanishes:
$$
T^{\mu}_{\mu (EM)} = g_{\mu\nu}T^{\mu\nu}_{EM} = F^{\mu\alpha}F_{\mu\alpha} - \frac{1}{4}(4)F_{\alpha\beta}F^{\alpha\beta} = 0.
$$
This result ($T^\mu_\mu=0$) is a consequence of the conformal invariance (scale invariance) of the classical Maxwell action in 4 dimensions. Since the Radion couples to the trace of the energy-momentum tensor ($\phi T^\mu_\mu$), one might naively conclude that the Radion does not interact with photons. However, this conformal invariance is broken at the quantum level. This phenomenon is known as the {trace anomaly} (or conformal anomaly)~\cite{Adler}.
To understand the origin of the Radion-photon coupling, we must look at the definition of the energy-momentum tensor in Quantum Field Theory. By definition, $T^{\mu\nu}$ is the response of the effective action $S_{eff}$ to a variation of the metric tensor $g_{\mu\nu}$:
$$
T^{\mu\nu} = \frac{2}{\sqrt{-g}} \frac{\delta S_{eff}}{\delta g_{\mu\nu}}.
$$
The trace of this tensor, $T^\mu_\mu = g_{\mu\nu}T^{\mu\nu}$, corresponds physically to the response of the action under a local conformal scaling of the metric, $g_{\mu\nu}(x) \to e^{2\omega(x)} g_{\mu\nu}(x)$. For an infinitesimal transformation ($\omega(x) \ll 1$), the variation of the metric is $\delta g_{\mu\nu} = 2\omega g_{\mu\nu}$. Substituting this into the definition of the tensor, the variation of the action becomes:
$$
\delta S_{eff} = \int d^4x \frac{\delta S_{eff}}{\delta g_{\mu\nu}} \delta g_{\mu\nu} = \int d^4x \left( \frac{\sqrt{-g}}{2} T^{\mu\nu} \right) (2\omega g_{\mu\nu}) = \int d^4x \sqrt{-g} \, \omega(x) \, T^\mu_\mu.
$$
If the theory were scale-invariant (conformal), the action would not change under rescaling ($\delta S_{eff} = 0$), implying $T^\mu_\mu = 0$. However, in the quantum theory, this symmetry is broken by the renormalization procedure. To define the theory, we must introduce a renormalization scale $\mu$ (an energy scale) to subtract divergences. The effective Lagrangian for the electromagnetic field, with the coupling constant factored out, takes the form
$$
\mathcal{L}_{eff} = -\frac{1}{4 e^2(\mu)} F_{\alpha\beta} F^{\alpha\beta} \sqrt{-g}.
$$
Here, we use the rescaled field variables ($A_\mu \to A_\mu/e$) so that the coupling $e(\mu)$ appears as a prefactor. A conformal transformation of the spacetime coordinates $x^\mu \to \lambda x^\mu$ is physically equivalent to rescaling the energy scale $\mu \to \mu/\lambda$. Therefore, the variation of the Lagrangian under a scale change is governed by the dependence of the coupling $e$ on the scale $\mu$. Mathematically, the trace is given by the renormalization group equation
$$
T^\mu_\mu = \mu \frac{\partial}{\partial \mu} \mathcal{L}_{eff} \bigg|_{\text{flat space}} = -\frac{1}{4} F_{\alpha\beta}F^{\alpha\beta} \left( \mu \frac{\partial}{\partial \mu} \frac{1}{e^2(\mu)} \right).
$$
We can evaluate the derivative term using the definition of the beta function, $\beta(e) \equiv \mu \frac{\partial e}{\partial \mu}$:
$$
\mu \frac{\partial}{\partial \mu} \left( e^{-2} \right) = -2 e^{-3} \left( \mu \frac{\partial e}{\partial \mu} \right) = -2 e^{-3} \beta(e).
$$
Substituting this result back into the trace expression:
$$
T^\mu_\mu = -\frac{1}{4} F^2 \left( -2 e^{-3} \beta(e) \right) = \frac{\beta(e)}{2e^3} (e^2 F^2_{canonical}).
$$
Restoring the canonical normalization for the fields (where the coupling is inside the covariant derivative, effectively removing the $e^2$ factor from our specific notation choice above), we arrive at the standard trace anomaly formula:
$$
T^\mu_\mu = \frac{\beta(e)}{2e} F_{\mu\nu} F^{\mu\nu}.
$$
This derivation proves that the non-zero trace is directly proportional to the beta function. Since the Radion field $\phi(x)$ couples to the trace of the energy-momentum tensor via $\frac{\phi}{\Lambda_r} T^\mu_\mu$, the interaction Lagrangian becomes:
$$
\mathcal{L}_{\phi\gamma\gamma} = \frac{\phi}{\Lambda_r} \left( \frac{\beta_{EM}(e)}{2e} F_{\mu\nu}F^{\mu\nu} \right).
$$
In Quantum Electrodynamics (QED), the one-loop beta function is $\beta_{EM}(e) = \frac{e^3}{12\pi^2}$. Injecting this expression into the Lagrangian yields:
$$
\mathcal{L}_{\phi\gamma\gamma} = \frac{\phi}{\Lambda_r} \frac{e^3}{24\pi^2 e} F_{\mu\nu}F^{\mu\nu} = \frac{\phi}{\Lambda_r} \frac{e^2}{24\pi^2} F_{\mu\nu}F^{\mu\nu}.
$$
Finally, using the fine-structure constant $\alpha_{EM} = \frac{e^2}{4\pi}$, we can rewrite the coefficient as $\frac{e^2}{24\pi^2} = \frac{\alpha_{EM}}{6\pi}$. Thus, the effective Lagrangian describing the interaction between the Radion and photons reads:
\be
\label{int2}
\mathcal{L}_{\phi\gamma\gamma} = \frac{\alpha_{EM}}{6\pi \Lambda_r} \phi F_{\mu\nu}F^{\mu\nu}.
\ee
This demonstrates that the Radion-photon coupling is a purely quantum mechanical effect generated by the breaking of scale invariance (the loop corrections).

\subsection{Constraints from Gravity Experiments}

The coupling of the Radion to the trace of the energy-momentum tensor implies that it mediates a "fifth force" roughly proportional to mass. It is crucial to check if such a force is compatible with experimental tests of General Relativity. We start with the interaction Lagrangian density derived previously:
$$
    \mathcal{L}_{\text{int}} = \frac{\phi}{\Lambda_r} T^\mu_\mu.
$$
For non-relativistic matter, the trace is dominated by the mass density, $T^\mu_\mu \approx \rho - 3p \simeq \rho$. Consider a point mass $M$ located at the origin. The scalar Radion field obeys the static Klein-Gordon equation with this source term:
$$
    (\nabla^2 - m_\phi^2) \phi = \frac{1}{\Lambda_r} T^\mu_\mu \approx \frac{M}{\Lambda_r} \delta^{(3)}(\mathbf{r}).
$$
The solution is the well-known Yukawa profile:
$$
    \phi(r) = - \frac{M}{4\pi \Lambda_r} \frac{e^{-m_\phi r}}{r}.
$$
The potential energy experienced by a second test mass $m$ due to the exchange of this field is $V_\phi(r) = \int d^3\mathbf{r}' \frac{\phi(\mathbf{r}')}{\Lambda_r} (m \delta^{(3)}(\mathbf{r}' - \mathbf{r}))$, which yields:
$$
    V_\phi(r) = - \frac{m}{\Lambda_r} \phi(r) = - \frac{M m}{4\pi \Lambda_r^2} \frac{e^{-m_\phi r}}{r}.
$$
The total static potential between the two masses, including standard gravity, is:
$$
    V(r) = V_G(r) + V_\phi(r) = -G \frac{M m}{r} \left( 1 + \alpha \, e^{-r/\lambda_r} \right),
$$
where $\lambda_r = 1/m_\phi$ is the Compton wavelength (interaction range) of the Radion, and the strength parameter $\alpha$ is given by:
$$
\alpha = \frac{1}{4\pi G \Lambda_r^2} = 2 \left( \frac{M_{Pl}}{\Lambda_r} \right)^2.
$$
For a typical symmetry breaking scale $\Lambda_r \sim 1$ TeV, the coupling strength is enormous, $\alpha \sim 10^{30}$. The phenomenological viability of the Radion therefore depends entirely on its mass $m_\phi$:

\begin{itemize}
    \item {Light Radion ($m_\phi \ll$ eV):} If the Radion were very light, the range $\lambda_r$ would be macroscopic (micrometers to meters). Torsion balance experiments (such as E\"{o}t-Wash) have strictly tested the inverse-square law at these distances. A modification with strength $\alpha \sim 10^{30}$ is completely excluded by these laboratory experiments. This forbids the Radion from being a light, long-range field.

    \item {Heavy Radion ($m_\phi >$ GeV):} For the mass range relevant to collider physics (GeV to TeV), the range of the force is sub-nuclear ($\lambda_r < 10^{-15}$ m). Macroscopic gravity experiments, which measure forces at separations $r \gg \lambda_r$, are insensitive to this Yukawa term because the exponential factor $e^{-r/\lambda_r}$ vanishes.
\end{itemize}

Consequently, while gravity experiments rule out a massless 
or light Radion solution to the hierarchy problem, they allow for a heavy, stabilized Radion. Theoretical mechanisms required to stabilize the extra dimension, such as the Goldberger-Wise mechanism \cite{Goldberger:1999uk}, typically generate a mass for the Radion that is suppressed relative to the scale $\Lambda_r$ only by a modest factor ($m_\phi \sim \Lambda_r / \text{few tens}$). Given that $\Lambda_r$ must be in the TeV range to solve the hierarchy problem, a Radion mass in the range of hundreds of GeV to a few TeV is a natural prediction of the theory. This provides the primary motivation for high-energy collider searches: the LHC is the only laboratory capable of probing this "short-range gravity" by directly producing the heavy Radion resonance.

\subsection{Coupling to the Higgs Boson and Kinetic Mixing}

Finally, we consider the coupling of the Radion to the Higgs scalar field $H$. This sector is particularly rich because it generates both mass mixing and kinetic mixing between the Radion and the physical Higgs boson.

\subsubsection{Standard Coupling via the Trace}
\label{sec:higgs_coupling}

We start with the Standard Model Higgs Lagrangian on the TeV-brane, given by:
\be
\label{eq:higgs_lag}
\mathcal{L}_{\text{Higgs}} = (D_\mu H)^\dagger (D^\mu H) - V(H),
\ee
where $D_\mu$ is the covariant derivative and the potential $V(H)$ responsible for spontaneous symmetry breaking is:
\be
V(H) = \lambda \left( |H|^2 - \frac{v^2}{2} \right)^2.
\ee
Here, $v \approx 246$ GeV is the vacuum expectation value (VEV) of the Higgs field. To find the coupling to the Radion, we compute the trace of the energy-momentum tensor $T^\mu_\mu$. The general expression for the energy-momentum tensor of a scalar field is:
$$
T_{\mu\nu} = (D_\mu H)^\dagger (D_\nu H) + (D_\nu H)^\dagger (D_\mu H) - g_{\mu\nu} \mathcal{L}_{\text{Higgs}}.
$$
Taking the trace with the metric $g^{\mu\nu}$ (where $g^{\mu\nu}g_{\mu\nu} = 4$ in 4 dimensions):
$$
T^\mu_\mu = 2 (D_\mu H)^\dagger (D^\mu H) - 4 \left[ (D_\mu H)^\dagger (D^\mu H) - V(H) \right],
$$
which simplifies to:
\be
T^\mu_\mu = -2 (D_\mu H)^\dagger (D^\mu H) + 4 V(H).
\ee
At first glance, this expression seems to depend on both the kinetic and potential terms. However, we can simplify it by using the classical Equation of Motion (EOM) for the Higgs field:
$$
D_\mu D^\mu H = - \frac{\partial V}{\partial H^\dagger} = - 2\lambda \left( |H|^2 - \frac{v^2}{2} \right) H.
$$
Using the identity $(D_\mu H)^\dagger (D^\mu H) \approx - H^\dagger (D_\mu D^\mu H)$ (valid up to total derivatives which vanish in the action), we can rewrite the kinetic term in the trace as:
$$
-2 (D_\mu H)^\dagger (D^\mu H) \approx 2 H^\dagger \left[ - 2\lambda \left( |H|^2 - \frac{v^2}{2} \right) H \right] = -4\lambda |H|^4 + 2\lambda v^2 |H|^2.
$$
Expanding the potential term $4V(H) = 4\lambda (|H|^4 - v^2 |H|^2 + v^4/4)$, we sum the contributions to find the trace:
$$
T^\mu_\mu = \left( -4\lambda |H|^4 + 2\lambda v^2 |H|^2 \right) + \left( 4\lambda |H|^4 - 4\lambda v^2 |H|^2 + \lambda v^4 \right).
$$
The quartic terms exactly cancel, leaving only the quadratic mass terms (and a constant vacuum energy term):
\be
\label{eq:trace_higgs}
T^\mu_\mu = - 2\lambda v^2 |H|^2 + \text{const}.
\ee
This result is consistent with the general theorem that the trace of the energy-momentum tensor is proportional to the scale-invariance breaking parameter of the theory. In the Higgs sector, the scale symmetry is broken by the tachyonic mass parameter $\mu^2 = -\lambda v^2$. To find the physical couplings, we expand the Higgs doublet around its vacuum state in the unitary gauge:
$$
H(x) = \begin{pmatrix} 0 \\ \frac{v + h(x)}{\sqrt{2}} \end{pmatrix},
$$
where $h(x)$ is the physical Higgs boson field. Substituting this into the trace:
$$
|H|^2 = \frac{1}{2}(v + h)^2 = \frac{1}{2}v^2 + vh + \frac{1}{2}h^2.
$$
The trace becomes:
$$
T^\mu_\mu = - \lambda v^2 (v^2 + 2vh + h^2) = - \lambda v^4 - 2\lambda v^3 h - \lambda v^2 h^2.
$$
Recalling that the physical Higgs mass is given by $m_h^2 = 2\lambda v^2$, we can rewrite the trace in terms of the physical mass:
\be
T^\mu_\mu = - \frac{m_h^2}{2} (2vh + h^2) + \text{const}.
\ee
Finally, utilizing the general interaction Lagrangian $\mathcal{L}_{int} = \frac{\phi}{\Lambda_r} T^\mu_\mu$, we obtain the explicit coupling terms between the Radion $\phi$ and the physical Higgs boson $h$:
\be
\mathcal{L}_{\phi h} = - \frac{m_h^2}{\Lambda_r} \left( v \phi h + \frac{1}{2} \phi h^2 \right).
\ee
This result contains two crucial interaction terms:
\begin{itemize}
    \item {Radion-Higgs Mixing ($\phi h$):} The term proportional to $v \phi h$ represents a kinetic mixing between the Radion and the Higgs boson. This implies that the physical mass eigenstates are actually mixtures of the geometry fluctuation ($\phi$) and the electroweak symmetry breaking scalar ($h$).
    \item {Radion-Higgs Interaction ($\phi h^2$):} The second term describes the decay of a heavy Radion into two Higgs bosons ($\phi \to hh$) or the production of a Radion in association with a Higgs.
\end{itemize}

\subsubsection{Higgs-Curvature Kinetic Mixing}
In addition to the standard coupling, general covariance allows for a non-minimal coupling between the Higgs field and the spacetime curvature on the brane. This is described by the action \cite{Giudice}:
\be
S_\xi = - \xi \int d^4x \sqrt{-g} \ R(g) H^\dagger H,
\ee
where $R(g)$ is the 4D Ricci scalar of the induced metric and $\xi$ is a dimensionless mixing parameter. In the linearized approximation, the induced metric is $g_{\mu\nu} \approx \eta_{\mu\nu} + 2\frac{\phi}{\Lambda_r}\eta_{\mu\nu}$. Under such a conformal rescaling, the Ricci scalar transforms as:
$$
R \approx - \frac{6}{\Lambda_r} \Box \phi.
$$
Substituting this into the action $S_\xi$ and retaining terms linear in the fields:
$$
S_\xi \approx - \xi \int d^4x \left( - \frac{6}{\Lambda_r} \Box \phi \right) \left( \frac{v^2}{2} + vh + \dots \right).
$$
We focus on the mixing term between $\phi$ and $h$:
$$
S_{\phi h}^{(mix)} = 6\xi \frac{v}{\Lambda_r} \int d^4x (\Box \phi) h.
$$
Integrating by parts (and discarding the boundary term), we transfer the derivative:
$$
\int d^4x (\Box \phi) h = \int d^4x (\partial_\mu \partial^\mu \phi) h = - \int d^4x \partial_\mu \phi \partial^\mu h.
$$
Thus, we obtain a kinetic mixing term in the Lagrangian:
\be
\label{kin_mix}
\mathcal{L}_{\phi h}^{(kin)} = - 6\xi \frac{v}{\Lambda_r} \partial_\mu \phi \partial^\mu h.
\ee
This term is crucial for phenomenology. It implies that the interaction eigenstates (the Radion $\phi$ and Higgs $h$) are not the physical mass eigenstates. To interpret experimental results, one must diagonalize the kinetic and mass terms simultaneously, leading to physical states that are admixtures of the Radion and the Standard Model Higgs.

\subsection{Linearization and Diagonalization of the Higgs-Radion Sector}

We can now construct the complete effective Lagrangian for the scalar sector by collecting the kinetic and mass terms derived in the previous sections. Combining the SM Higgs Lagrangian (Eq.~\ref{eq:higgs_lag}), the Radion Lagrangian (Eq.~\ref{kin1}), and the mixing terms induced by the trace anomaly (Eq.~\ref{eq:trace_higgs}) and the non-minimal curvature coupling (Eq.~\ref{kin_mix}), the total action reads
$$S = \int d^4x \sqrt{-g} \left[ \underbrace{g^{\mu\nu} (D_\mu H)^\dagger (D_\nu H)}_{\text{Kinetic Term}} - \underbrace{\lambda(|H|^2 - v^2)^2}_{\text{Potential Term } V(H)} - \underbrace{\xi R H^\dagger H}_{\text{Non-minimal Coupling}} \right]$$
or
\be
S_{tot} = \int d^4x \left[ \mathcal{L}_{kin} - \mathcal{L}_{mass} \right].
\ee
Using the notation $\gamma \equiv v/\Lambda_r$ introduced in Section 5, we linearize the fields around their vacuum expectation values. Retaining only terms bilinear in the fluctuation fields $h$ and $\phi$, the kinetic and mass sectors can be written in matrix form. The kinetic Lagrangian density, including the mixing derived in Eq.~(\ref{kin_mix}), is:
\be
\mathcal{L}_{kin}^{(2)} = \frac{1}{2} 
\begin{pmatrix} \partial_\mu h & \partial_\mu \phi \end{pmatrix} 
\begin{pmatrix} 
1 & -6\xi\gamma \\ 
-6\xi\gamma & 1 
\end{pmatrix} 
\begin{pmatrix} \partial^\mu h \\ \partial^\mu \phi \end{pmatrix}.
\ee
The off-diagonal term $-6\xi\gamma$ represents the kinetic mixing parameter.
To ensure the kinetic energy is positive definite, we require the determinant to be positive, $1 - (6\xi\gamma)^2 > 0$.
The mass Lagrangian density, combining the standard masses and the trace anomaly mixing from Eq.~(\ref{eq:trace_higgs}), is:
\be
\mathcal{L}_{mass}^{(2)} = \frac{1}{2} 
\begin{pmatrix} h & \phi \end{pmatrix} 
\hat{M}^2 
\begin{pmatrix} h \\ \phi \end{pmatrix} 
= \frac{1}{2} 
\begin{pmatrix} h & \phi \end{pmatrix} 
\begin{pmatrix} 
m_h^2 & \gamma m_h^2 \\ 
\gamma m_h^2 & m_\phi^2 
\end{pmatrix} 
\begin{pmatrix} h \\ \phi \end{pmatrix}.
\ee
The physical mass eigenstates are found by diagonalizing this system. This requires a two-step procedure: first to canonically normalize the kinetic term, and second to diagonalize the resulting mass matrix.
We define a renormalized field basis $(h', \phi')$ to eliminate the cross-derivative term and restore canonical normalization. A correct transformation that achieves this is:
\be
h = \frac{h'}{Z_k}, \quad \phi = \phi' + \frac{6\xi\gamma}{Z_k} h',
\ee
where $Z_{k} = \sqrt{1 - (6\xi\gamma)^2}$ is the kinetic renormalization factor. Substituting these into $\mathcal{L}_{kin}^{(2)}$, we recover the canonical form $\frac{1}{2}(\partial h')^2 + \frac{1}{2}(\partial \phi')^2$.
Applying the kinetic transformation to the mass sector induces new terms in the mass matrix. The potential becomes:
\be
\label{mat_mass}
V(h', \phi') = \frac{1}{2} 
\begin{pmatrix} h' & \phi' \end{pmatrix} 
\mathcal{M}^2 
\begin{pmatrix} h' \\ \phi' \end{pmatrix},
\ee
with the entries of the new squared-mass matrix $\mathcal{M}^2$:
\begin{align}
\mathcal{M}^2_{11} &= \frac{1}{Z_k^2} \left( m_h^2 (1+12\xi\gamma^2) + m_\phi^2 (6\xi\gamma)^2 \right), \\
\mathcal{M}^2_{22} &= m_\phi^2, \\
\mathcal{M}^2_{12} &= \frac{1}{Z_k} \left( \gamma m_h^2 + 6\xi\gamma m_\phi^2 \right).
\end{align}
Finally, this symmetric matrix is diagonalized via an orthogonal rotation by an angle $\theta$:
\be
\begin{pmatrix} h_{phys} \\ \tilde{\phi} \end{pmatrix} = 
\begin{pmatrix} 
\cos\theta & \sin\theta \\ 
-\sin\theta & \cos\theta 
\end{pmatrix} 
\begin{pmatrix} h' \\ \phi' \end{pmatrix}.
\ee
The physical masses squared, $m_{h_{phys}}^2$ and $m_{\tilde{\phi}}^2$, are the eigenvalues of $\mathcal{M}^2$. The mixing angle $\theta$ is given by:
\be
\label{theta_calc}
\tan 2\theta = \frac{2 \mathcal{M}^2_{12}}{\mathcal{M}^2_{11} - \mathcal{M}^2_{22}}.
\ee
This angle dictates the modification of the couplings to the Standard Model particles presented in Eq.~(\ref{eq:phys_coupling_final}) and used in the simulation setup in Section 5.

\subsection{Comparison with the Axion-Photon Coupling}

It is insightful to compare the effective Lagrangian derived in Eq.~(\ref{int2}) 
for the Radion with the standard coupling of an Axion-Like Particle (ALP) to photons. Both the Radion ($\phi$) and the axion ($a$) can decay into two photons and are often searched for in similar experimental channels (such as light-by-light scattering or diphoton resonances). However, their couplings arise from different physical mechanisms and possess distinct symmetry properties. The Radion-photon interaction, originating from the trace anomaly, couples a scalar field ($J^{P}=0^{+}$) to the electromagnetic invariant $F_{\mu\nu}F^{\mu\nu}$:
\be
\label{Radion_coupling}
\mathcal{L}_{\phi\gamma\gamma} = \frac{\alpha_{EM}}{6 \pi \Lambda_{r}} \phi F_{\mu\nu}F^{\mu\nu}.
\ee
In contrast, an axion or ALP is a pseudo-scalar field ($J^{P}=0^{-}$). Its coupling to photons arises from the chiral anomaly (connected to the topological term $F_{\mu\nu}\tilde{F}^{\mu\nu}$) and takes the form:
\be
\label{axion_coupling}
\mathcal{L}_{a\gamma\gamma} = \frac{g_{a\gamma\gamma}}{4} a F_{\mu\nu}\tilde{F}^{\mu\nu}=\frac{1}{4\Lambda_a} a F_{\mu\nu}\tilde{F}^{\mu\nu}, 
\ee
where $\tilde{F}^{\mu\nu} = \frac{1}{2} \epsilon^{\mu\nu\alpha\beta} F_{\alpha\beta}$ is the dual field strength tensor.

\section{Scattering of light-by-light at the LHC}

\subsection{Introduction}

The scattering of light-by-light $\gamma\gamma \rightarrow \gamma\gamma$
can be probed in proton-proton, proton-ion or ion-ion collisions, where the initial state photons corresponding to electromagnetic fields (EM) produced by the ultra-relativistic incident hadrons (protons or ions). These interactions occur in ultra-peripheral collisions (UPC), where the two incident hadrons do not collide centrally but are separated in the transverse direction by at least the sum of their radii. In the SM, the scattering of light-by-light is induced at one-loop at leading order (quark, charged leptons and $W$ boson contributions). The signature of these reactions is the presence of two photons and no additional activity in the central detector. The outgoing hadrons (protons or ions) escape into the beam pipe. In the following, we focus on high-mass searches in proton-proton collisions, where the exclusive signal can be identified using forward proton spectrometers. Radions of masses above a few GeV  can then be probed at the LHC, as a deviation on the elastic scattering of light-by-light through the reaction 
\begin{equation}
\gamma \gamma \rightarrow \phi \rightarrow \gamma \gamma,
\label{toto2}
\end{equation}
thus modifying the di-photon invariant mass distribution. Indeed, a Radion of mass $m_\phi$ would appear as a resonance in the distribution of the invariant mass of the photon pair.

\subsection{Influence of Higgs-Radion Mixing on the Diphoton Signal}

The search for the Radion via light-by-light scattering (Reaction~(\ref{toto2})) relies on the effective Radion-photon coupling derived in Eq.~(\ref{int2}). However, the presence of the Radion-Higgs mixing term derived in Eq.~(\ref{kin_mix}) significantly modifies this phenomenology. The mixing term $\mathcal{L}_{\phi h} \propto v \phi h$ implies that the interaction eigenstates $\phi$ (geometry fluctuation) and $h$ (SM Higgs) are not the physical mass eigenstates. Upon diagonalization, the physical Radion state $\tilde{\phi}$ acquires a small admixture of the SM Higgs component:
\be
\tilde{\phi} \approx \cos\theta \, \phi - \sin\theta \, h,
\ee
where $\theta$ is the mixing angle governed by the parameter $\xi$ and the mass hierarchy. This mixing induces a modification of the effective coupling to photons. The Standard Model Higgs boson $h$ couples to photons primarily through $W$-boson and top-quark loops. Therefore, the coupling of the physical Radion $\tilde{\phi}$ to the diphoton final state becomes a coherent sum of two contributions:
\begin{itemize}
    \item The {trace anomaly} contribution (pure Radion part): proportional to $\frac{\alpha_{EM}}{6\pi \Lambda_r}$.
    \item The {Higgs-like} contribution (mixing part): proportional to $\sin\theta \times \frac{\alpha_{EM}}{8\pi v} \mathcal{A}_{SM}$.
\end{itemize}
Mathematically, the effective Lagrangian for the physical state becomes:
\be
\label{eq:phys_coupling_def}
\mathcal{L}_{\tilde{\phi}\gamma\gamma} = \left( \frac{\alpha_{EM}}{6\pi \Lambda_r} \cos\theta - \frac{\alpha_{EM}}{8\pi v}\mathcal{A}_{SM} \sin\theta \right) \tilde{\phi} F_{\mu\nu}F^{\mu\nu}.
\ee

Let us find Eq.~(\ref{eq:phys_coupling_def}) explicitely. To derive the coupling of the physical mass eigenstates to photons, we must start with the interaction eigenstates in the unmixed Lagrangian and apply the rotation induced by the Higgs-Radion mixing.
Before diagonalization, the theory contains two distinct scalar fields: the unmixed Radion $\phi_0$ (representing purely geometric fluctuations) and the unmixed Higgs boson $h_0$ (the $SU(2)$ doublet component). Their couplings to the photon field strength tensor $F_{\mu\nu}$ arise from different sources:
\begin{enumerate}
    \item {Radion Coupling ($\phi_0$):} As derived in Section 3.2, the Radion couples via the trace anomaly of the energy-momentum tensor:
    \be
    \mathcal{L}_{\phi_0 \gamma\gamma} = \frac{\alpha_{EM}}{6\pi \Lambda_r} \phi_0 F_{\mu\nu}F^{\mu\nu}.
\ee
    \item {Higgs Coupling ($h_0$):} The Standard Model Higgs couples to photons primarily through $W$-boson and top-quark loops. The effective Lagrangian is:
    \be
    \mathcal{L}_{h_0 \gamma\gamma} = \frac{\alpha_{EM}}{8\pi v} \mathcal{A}_{SM} h_0 F_{\mu\nu}F^{\mu\nu},
    \ee
    where $v \approx 246$ GeV is the electroweak vacuum expectation value and $\mathcal{A}_{SM} \approx -6.5$ is the dimensionless loop function.
\end{enumerate}

Then, we can move to the physical basis of mass eigenstates, denoted by $\tilde{\phi}$ (the physical Radion) and $\tilde{h}$ (the physical Higgs). For the component of the interaction eigenstates contributing to the physical Radion $\tilde{\phi}$, the projection is:
\begin{align}
\phi_0 &\supset \cos\theta \, \tilde{\phi} \\
h_0    &\supset -\sin\theta \, \tilde{\phi}
\end{align}
where we have adopted the sign convention that yields $h_0 = \cos\theta \tilde{h} - \sin\theta \tilde{\phi}$.
Now we substitute these projections into the total interaction Lagrangian $\mathcal{L}_{int} = \mathcal{L}_{\phi_0 \gamma\gamma} + \mathcal{L}_{h_0 \gamma\gamma}$ to isolate the coupling to the physical Radion $\tilde{\phi}$:
$$
\mathcal{L}_{\tilde{\phi}\gamma\gamma} = \left[ \frac{\alpha_{EM}}{6\pi \Lambda_r} (\cos\theta \, \tilde{\phi}) + \frac{\alpha_{EM}}{8\pi v} \mathcal{A}_{SM} (-\sin\theta \, \tilde{\phi}) \right] F_{\mu\nu}F^{\mu\nu}.
$$
Factoring out the common terms, we obtain the final expression for the effective coupling:
\be
\label{eq:phys_coupling_final}
\mathcal{L}_{\tilde{\phi}\gamma\gamma} = \frac{\alpha_{EM}}{6\pi \Lambda_r} \left( \cos\theta - \frac{3}{4}\mathcal{A}_{SM} \left(\frac{\Lambda_r}{v}\right) \sin\theta \right) \tilde{\phi} F_{\mu\nu}F^{\mu\nu}.
\ee
This confirms the interference between the direct gravitational coupling and the induced Higgs-like coupling.

\subsubsection{Phenomenological Influence of the Mixing Parameter $\xi$}

\begin{figure}[htbp]
\centering
\includegraphics[width=1.0\textwidth]{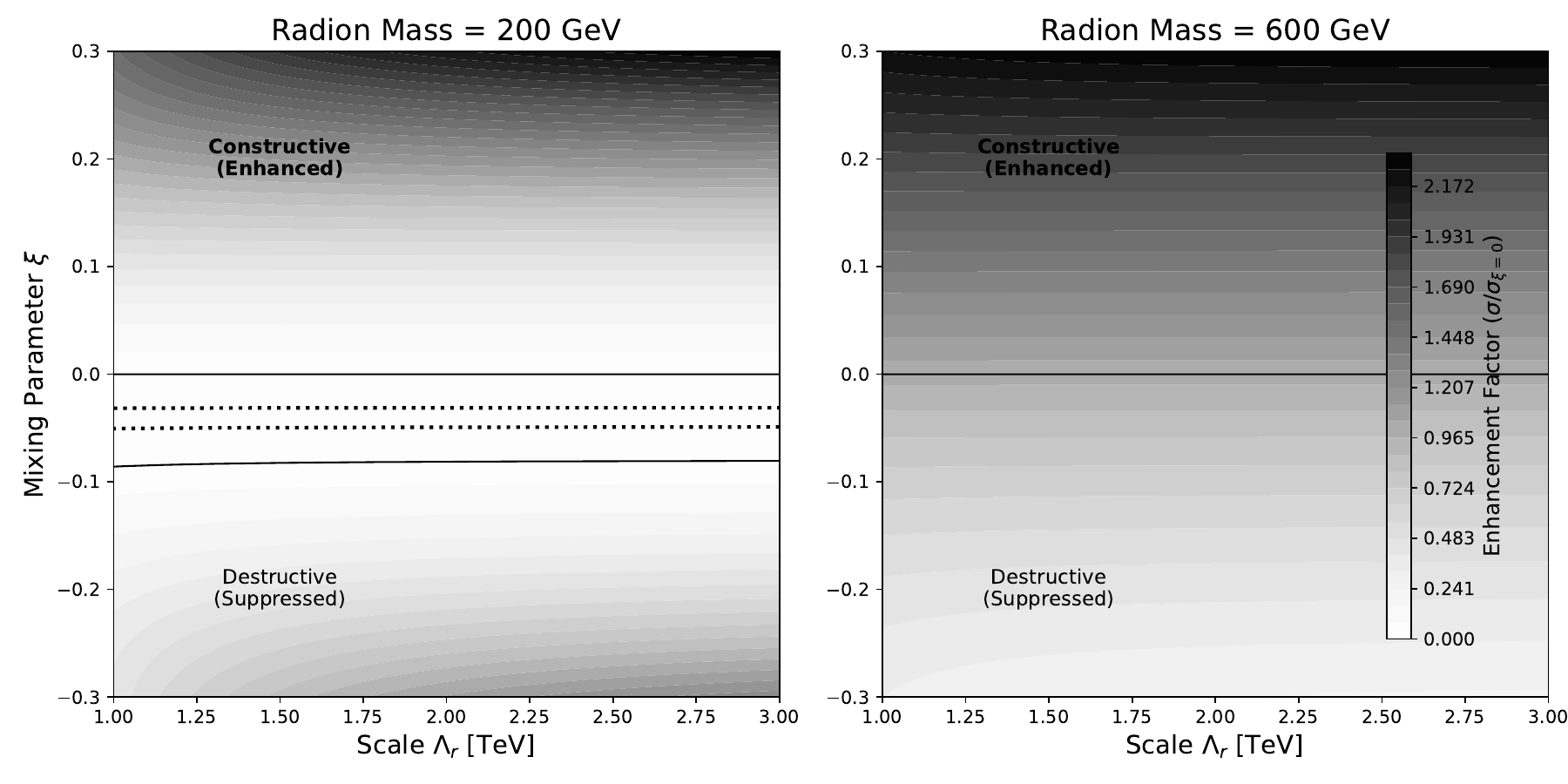}
\caption{The interference landscape for the Radion (Left: $m_\phi=200$ GeV, Right: $m_\phi=600$ GeV) in the $(\Lambda_r, \xi)$ plane. The greyscale indicates the enhancement factor relative to the pure curvature-induced prediction. Darker regions indicate constructive interference ($\xi > 0$), while lighter regions indicate destructive interference. The dotted line marks the "photophobic" region where the signal is suppressed.}
\label{fig:interference}
\end{figure}

The parameter $\xi$ (originating from the non-minimal curvature coupling $\xi R H^\dagger H$) is the driving force behind the mixing angle $\theta$. Its influence on the diphoton signal is profound because it controls the interference between the two terms in Eq.~(\ref{eq:phys_coupling_def}).
\begin{itemize}
    \item {Relationship between $\xi$ and $\theta$:} 
    In the limit of small mixing ($v \ll \Lambda_r$), the mixing angle is approximately proportional to $\xi$ and the Radion mass squared:
    \begin{equation}
    \label{eq:mixing_approx}
    \tan 2\theta \approx 12\xi \left(\frac{v}{\Lambda_r}\right) \frac{m_{\phi}^2}{m_{\phi}^2 - m_{h}^2}.
\end{equation}
    Assuming $m_\phi > m_h$, the sign of the mixing angle $\theta$ follows the sign of $\xi$.

{\it Let us make a note on basis dependence:} 
    In the literature (e.g. Ref. \cite{Csaki:2007ns}), a formula proportional to $m_h^2$ (rather than $m_\phi^2$) is often encountered for the mixing angle. This difference arises from the choice of field basis during the kinetic diagonalization. If one shifts the Higgs field ($h \to h' + \delta \phi$) to eliminate mixing, the angle scales with $m_h^2$. However, if one shifts the Radion field ($\phi \to \phi' + \delta h$), as we did in Step 1 of Section 3.5, the angle scales with $m_\phi^2$. Since our effective couplings in Eq.~(\ref{eq:phys_coupling_final}) are derived in the latter basis, consistency requires the use of the formula above. This ensures the correct non-decoupling behavior for large $\xi$.

\item {Interference Patterns:}
    The effective coupling squared (which determines the cross-section) is proportional to:
    $$ |g_{eff}|^2 \propto \left|
\cos\theta - \frac{3}{4} \mathcal{A}_{SM} \left(\frac{\Lambda_r}{v}\right) \sin\theta \right|^2. $$
    Since the SM loop function is negative ($\mathcal{A}_{SM} \approx -6.5$), the term $-\mathcal{A}_{SM}$ is positive. Therefore, the relative sign of $\theta$ determines the nature of the interference:
    
    \begin{enumerate}
        \item {Destructive Interference ($\xi < 0$):} 
        If $\xi$ is negative, $\sin\theta$ is negative. The Higgs-loop term subtracts from the trace anomaly term. Specifically, if the condition $\tan\theta \approx \frac{4}{3} \frac{v}{\Lambda_r \mathcal{A}_{SM}}$ is met, the effective coupling vanishes. In this `photophobic' region (shown as the black line in Fig.~\ref{fig:interference}), the Radion becomes invisible to $\gamma\gamma$ searches.

        \item {Constructive Interference ($\xi > 0$):} 
        If $\xi$ is positive, $\sin\theta$ is positive. The Higgs-loop contribution adds constructively to the trace anomaly term. Because the ratio $\Lambda_r/v$ is large (typically $\mathcal{O}(10)$), even a small mixing $\xi$ results in a dominant Higgs-like contribution. This leads to a significant enhancement of the diphoton signal (the red region in Fig.~\ref{fig:interference}).
    \end{enumerate}
\end{itemize}

In {conclusion,} the search for the Radion in light-by-light scattering is not merely a probe of the scale $\Lambda_r$, but a sensitive probe of the mixing parameter $\xi$. Experimental limits must therefore be interpreted as exclusion contours in the $(\Lambda_r, \xi)$ plane.

\section{Simulation and analysis}
\label{sec:simulation}

\subsection{Theoretical Model and Event Generation}

In this work, we adopt an effective field theory (EFT) approach to model the heavy resonance production in ultra-peripheral collisions. The simulation framework integrates the \textsc{gamma-UPC} code~\cite{Shao:2022cyd} to handle the photon fluxes and \textsc{MadGraph5\_aMC@NLO}~\cite{Alwall:2014hca, Frederix:2018nkq} for the matrix element calculations. The effective Lagrangians describing the Radion and ALP interactions, as detailed in Section 3, are implemented via Universal Feynman Output (UFO) models~\cite{Degrande:2011ua, Darme:2023jdn}. 
We generate signal events for the exclusive production process $pp \to p(\gamma\gamma \to X)p$, where $X$ is either the Radion ($\phi$) or the Axion-Like Particle ($a$), decaying into a diphoton final state. The simulation is performed at a center-of-mass energy of $\sqrt{s} = 14$ TeV, consistent with the LHC Run 3 and High-Luminosity (HL-LHC) configurations.

\subsection{Event Selection for Proton-Proton Collisions}

As high-luminosity proton-proton collisions suffer from significant pileup interactions,  the analysis relies on the detection of the intact outgoing protons using forward spectrometer detectors (such as the ATLAS AFP~\cite{AFP} or CMS CT-PPS~\cite{CTPPS}),
in order to select clean exclusive photon-fusion events. Based on recent LHC searches for high-mass exclusive resonances~\cite{Baldenegro:2018hng,Schoeffel:2020svx}, our analysis targets the high-mass region ($m_X > 150$ GeV). We apply the following selection criteria to the simulated signal samples, mirroring the experimental acceptance:

\begin{itemize}
    \item {Photon Kinematics:} We require two photons with high transverse momentum, $p_T^{\gamma} > 40$ GeV, within the detector acceptance $|\eta^{\gamma}| < 2.37$.
    \item {Diphoton System:} The invariant mass of the diphoton pair must satisfy $m_{\gamma\gamma} > 150$ GeV. To suppress non-exclusive backgrounds, the system must be back-to-back in the transverse plane, enforced by an acoplanarity requirement $A_{\phi}^{\gamma\gamma} = 1 - |\Delta\phi_{\gamma\gamma}|/\pi < 0.01$.
    \item {Proton Tagging:} To ensure the event is exclusive, both outgoing protons must be within the acceptance of the forward detectors. This is modeled by requiring the fractional momentum loss of the protons, $\xi = 1 - E'_p/E_{beam}$, to be within the typical acceptance range $\xi \in [0.035, 0.08]$.
\end{itemize}

Table~\ref{tab:selection_pp} summarizes these selection criteria.

\begin{figure}[htbp]
\centering
\includegraphics[width=0.8\textwidth]{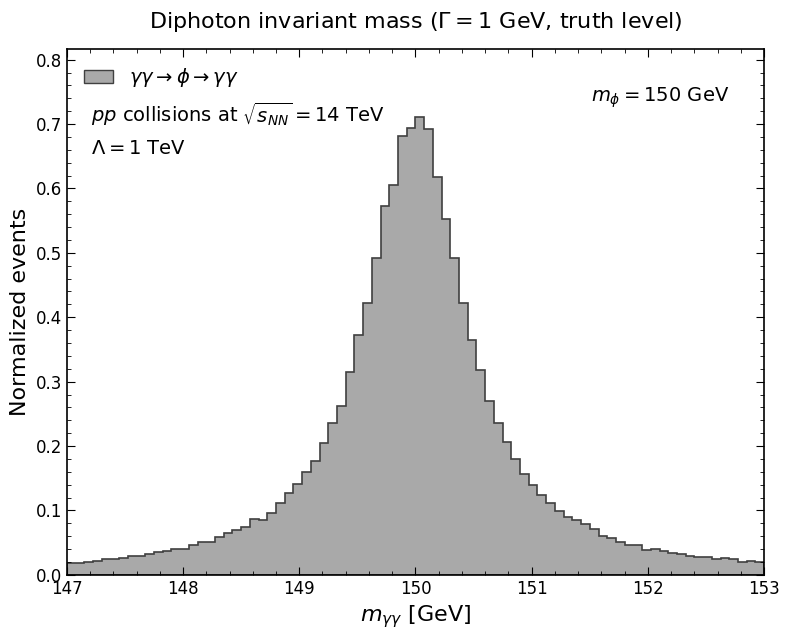}
\caption{Simulated diphoton invariant mass distribution (normalized to unity) for a Radion resonance ($m_{\phi} = 150$ GeV, $\Gamma = 1$ GeV) at truth level. This validation plot confirms the correct generation of the Breit-Wigner line shape by the Monte Carlo framework described in Section 5.1, within the kinematic requirements region defined in Table~\ref{tab:selection_pp}.}
\label{fig:truth_level_mass}
\end{figure}

In order to validate the simulation framework described in Section~\ref{sec:simulation} and ensure the correct implementation of the kinematic acceptance, we present in Figure~\ref{fig:truth_level_mass} the diphoton invariant mass distribution at truth level, normalized to unity. The plot displays the resonance structure for a benchmark Radion with $m_{\phi} = 150$ GeV and $\Gamma = 1$ GeV, confirming that the generated signal topology is well-defined and consistent with the theoretical line shape before the application of detector effects.

\begin{table}[h]
\centering
\begin{tabular}{|l|c|}
\hline
{Variable} & {Selection Criterion ($pp$ Analysis)} \\
\hline
\hline
$\sqrt{s}$ & 14 TeV \\
\hline
Photon $p_T$ & $> 40$ GeV \\
\hline
Photon Pseudorapidity & $|\eta| < 2.37$ \\
\hline
Diphoton Invariant Mass & $m_{\gamma\gamma} > 150$ GeV \\
\hline
Acoplanarity ($A_{\phi}$) & $< 0.01$ \\
\hline
Proton Tagging ($\xi$ range) & $0.035 < \xi < 0.08$ \\
\hline
\end{tabular}
\caption{Summary of the event selection criteria applied in the simulation for the proton-proton analysis, based on forward proton tagging acceptances.}
\label{tab:selection_pp}
\end{table}

\subsection{Limit Extraction Strategy}

\begin{figure}[htbp]
\centering
\includegraphics[width=0.8\textwidth]{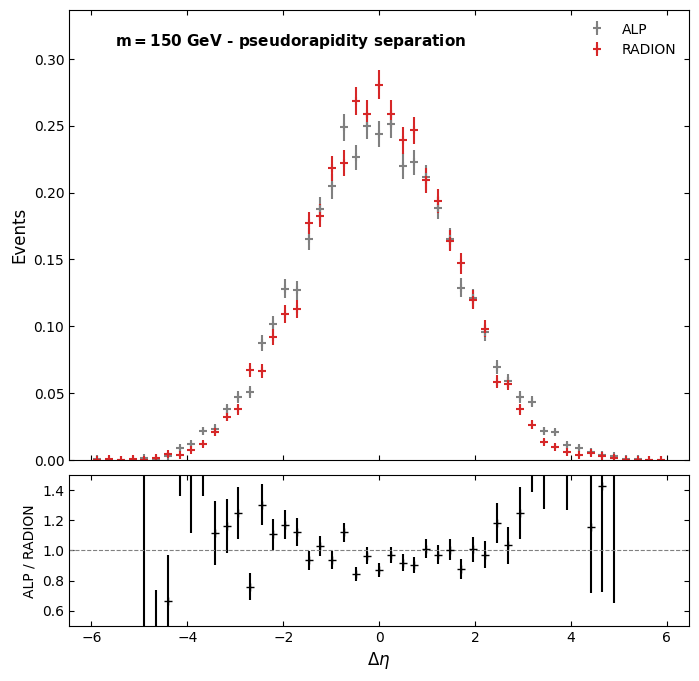}
\caption{ALP and Radion simulation distributions at truth level, both normalized to unity. The figure shows the comparison of the pseudorapidity separation $\Delta\eta$ between the two photons for ALP (grey) and Radion (red) signals at $m=150$ GeV. The close agreement supports the assumption that the selection efficiencies are comparable.}
\label{fig:efficiency_comparison}
\end{figure}

To derive exclusion limits on the Radion parameter space, we utilize the existing experimental constraints on Axion-Like Particles (ALPs) as a benchmark. Since both the ALP ($a$) and the Radion ($\phi$) are produced via photon fusion and decay into diphotons, their kinematic topologies are very similar. 
To justify translating the limits, we compare the kinematic distributions of both signals. Figure~\ref{fig:efficiency_comparison} shows the differential distribution of the pseudorapidity separation $\Delta\eta_{\gamma\gamma}$ between the decay photons for a resonance mass of 150 GeV. 
Although the ALP is a pseudo-scalar ($0^-$) and the Radion a scalar ($0^+$), the shape of the distributions at truth level, both normalized to unity, are nearly identical within the statistical uncertainties.
Given that the experimental acceptance is driven by the geometric coverage of the detector and the kinematic cuts on $p_T$ and $\eta$, this structural similarity allows us to assume that the selection efficiencies are comparable ($\mathcal{A}_{\phi} \approx \mathcal{A}_{a}$).
Consequently, we can translate an upper limit on the ALP coupling $g_{a\gamma}$ into a limit on the Radion coupling $g_{\phi\gamma}$. The signal yield $N$ scales as $N \propto \mathcal{L} \times \sigma \times \mathcal{A}$. Therefore, the limit conversion is given by:
\begin{equation}
    \label{eq:limit_conversion}
    \frac{\alpha_{EM}}{6\pi \Lambda_r^{limit}} = \frac{1}{4\Lambda_a^{limit}}\times \sqrt{ \frac{\sigma_{a}^{ref}}{\sigma_{\phi}^{ref}} },
\end{equation}
where $\sigma^{ref}$ denotes the reference production cross-section for a unit coupling. This relation allows us to re-interpret the grey exclusion region provided by LHC ALP searches~\cite{Baldenegro:2018hng,Schoeffel:2020svx} (shown in Section 6) directly into the $(\Lambda_r, \xi)$ plane for the Radion.

\section{Comparison with Axion-Like Particle Limits}
\label{sec:alp_comparison}

In this section, we interpret the sensitivity of light-by-light scattering searches in the context of the Radion, comparing it explicitly to the standard limits obtained for Axion-Like Particles (ALPs). While both particles induce a resonant deviation in the diphoton invariant mass spectrum, their coupling structures differ fundamentally, leading to significantly different constraints on their respective scales $\Lambda$.

\subsection{Analytical Comparison of Couplings}

The effective interaction of an ALP ($a$) with photons is traditionally parameterized by the dimension-5 operator:
\begin{equation}
    \mathcal{L}_{a\gamma\gamma} = \frac{1}{4\Lambda_a} a F_{\mu\nu}\tilde{F}^{\mu\nu},
\end{equation}
where $\Lambda_a$ represents the symmetry breaking scale of the ALP. In contrast, as derived in Eq.~(\ref{int2}), the Radion coupling arises from the trace anomaly and is loop-suppressed:
\begin{equation}
    \mathcal{L}_{\phi\gamma\gamma} = \frac{\alpha_{EM}}{6\pi \Lambda_r} \phi F_{\mu\nu}F^{\mu\nu}.
\end{equation}
Comparing the coefficients, the pure Radion coupling (in the absence of mixing) is suppressed relative to the ALP coupling by a factor of $\frac{2\alpha_{EM}}{3\pi} \approx 1.55 \times 10^{-3}$. Consequently, for an experimental analysis sensitive to a cross-section $\sigma_{lim}$, the derived limit on the Radion scale $\Lambda_r$ is naturally much weaker than that on the ALP scale $\Lambda_a$.
However, the Higgs-Radion mixing parameter $\xi$ drastically alters this picture. As shown in Eq.~(\ref{eq:phys_coupling_final}), the mixing induces a Higgs-like contribution that scales with $1/v$ rather than $1/\Lambda_r$. The effective coupling becomes:
\begin{equation}
    g_{eff}^{\phi} \approx \frac{\alpha_{EM}}{6\pi \Lambda_r} \left( 1 - \frac{3}{4} \mathcal{A}_{SM} \frac{\Lambda_r}{v} \theta \right),
\end{equation}
where $\mathcal{A}_{SM} \approx -6.5$ is the SM loop function. Using the small-angle approximation for the mixing angle derived in Eq.~(\ref{eq:mixing_approx}), $\theta \approx 6\xi (v/\Lambda_r)$, the interference term becomes independent of the scale $\Lambda_r$:
\begin{equation}
    \text{Enhancement Ratio} \equiv \frac{g_{eff}^{\phi}(\xi)}{g_{eff}^{\phi}(\xi=0)} \approx 1 - 4.5\xi\mathcal{A}_{SM}.
\end{equation}
For constructive interference ($\xi > 0$), this ratio provides a substantial enhancement to the signal.

\subsection{Comparison with ALP Cross-Sections at Fixed Scale}

It is instructive to quantify the suppression of the Radion signal relative to an Axion-Like Particle (ALP) when assuming the same fundamental scale, $\Lambda_r = \Lambda_a \equiv \Lambda$. Comparing the effective Lagrangians derived in Eqs.~(\ref{Radion_coupling}) and (\ref{axion_coupling}), the ratio of the scattering cross-sections is given by:
\begin{equation}
    \frac{\sigma_{Radion}(\Lambda)}{\sigma_{ALP}(\Lambda)} = \left|
\frac{g_{eff}^{\phi}}{g_{eff}^{a}} \right|^2 \approx \left( \frac{2\alpha_{EM}}{3\pi} \right)^2 \times \left( 1 - 4.5\xi\mathcal{A}_{SM} \right)^2.
\end{equation}

\begin{figure}[h]
\centering
\includegraphics[width=0.85\textwidth]{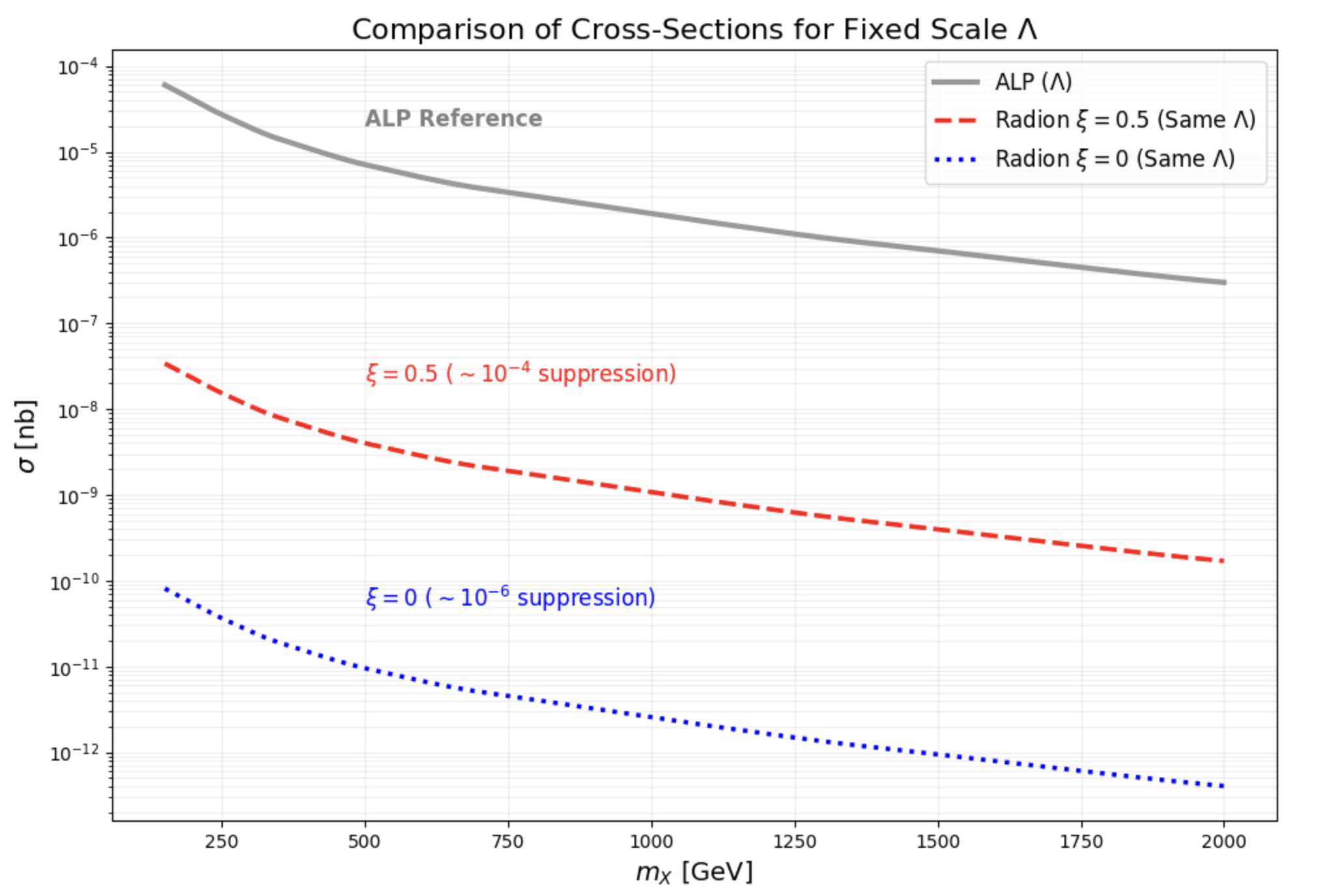}
\caption{Comparison of the light-by-light scattering cross-section $\sigma(\gamma\gamma \to X \to \gamma\gamma)$ for an ALP (grey solid line) and a Radion (colored dashed lines) as a function of mass, assuming a fixed scale $\Lambda = 1$ TeV for all models. The blue dotted line ($\xi=0$) shows the immense suppression of the pure trace anomaly coupling. The red dashed line ($\xi=0.5$) demonstrates the recovery of sensitivity due to Higgs-Radion mixing, though the rate remains suppressed relative to the ALP by a factor of $\sim 5 \times 10^{-4}$.}
\label{fig:xs_comparison}
\end{figure}

The term $\left( \frac{2\alpha_{EM}}{3\pi} \right)^2 \approx 2.40 \times 10^{-6}$ represents a severe suppression factor inherent to the loop-induced nature of the Radion coupling. While the Higgs-mixing term (dependent on $\xi$) enhances the Radion signal significantly, it does not fully overcome this suppression. This behavior is illustrated in Figure~\ref{fig:xs_comparison}, which compares the production cross-sections for an ALP and a Radion assuming a fixed benchmark scale of $\Lambda = 1$ TeV. The grey curve represents the ALP signal, while the red (dashed) and blue (dotted) curves represent the Radion signal with mixing ($\xi=0.5$) and without mixing ($\xi=0$), respectively. Even in the scenario of strong constructive interference ($\xi=0.5$), the Radion cross-section remains approximately $5.8 \times 10^{-4}$ times smaller than that of an ALP with the same scale.

Table~\ref{tab:xs_comparison_alp} summarizes the numerical suppression factors derived from these ratios. This explains why an experimental analysis sensitive to a $1$~TeV ALP can only exclude a Radion scale of order $\sim 24$~GeV.

\begin{table}[h]
\centering
\begin{tabular}{|c|c|c|c|}
\hline
{Mixing} & {Enhancement} & {Suppression Factor} & {Effective Ratio} \\
{Parameter} & {vs. Pure Radion} & {vs. ALP (at same $\Lambda$)} & {($\sigma_{Radion} / \sigma_{ALP}$)} \\
{($\xi$)} & {($\sigma(\xi)/\sigma(0)$)} & {(Loop factor only)} & {(Total)} \\
\hline
\hline
$\xi = 0$ & $1$ & $\approx 2.40 \times 10^{-6}$ & ${2.40 \times 10^{-6}}$ \\
\hline
$\xi = 0.2$ & $\approx 47$ & $\approx 2.40 \times 10^{-6}$ & ${1.13 \times 10^{-4}}$ \\
\hline
$\xi = 0.5$ & $\approx 243$ & $\approx 2.40 \times 10^{-6}$ & ${5.83 \times 10^{-4}}$ \\
\hline
\end{tabular}
\caption{Comparison of the Radion signal strength. The second column shows the enhancement due to mixing relative to the pure gravity case ($\xi=0$). The last column shows the total suppression of the Radion cross-section compared to an ALP cross-section for the same fixed scale $\Lambda$, combining the loop suppression $\left(\frac{2\alpha}{3\pi}\right)^2$ and the mixing enhancement.}
\label{tab:xs_comparison_alp}
\end{table}

\subsection{Signal Strength and Cross-Section Enhancement}

The phenomenological impact of the Higgs-Radion mixing is most clearly observed in the scattering cross-section, which scales with the square of the effective coupling, $\sigma \propto |g_{eff}|^2$. While the pure curvature-induced coupling ($\xi=0$) is suppressed by the trace anomaly factor, the constructive interference driven by the non-minimal coupling $\xi$ generates a rapid enhancement of the signal rate. Using the interference ratio derived in the previous section, the enhancement of the cross-section relative to the pure gravity prediction is given by:
\begin{equation}
    \frac{\sigma(\xi)}{\sigma(\xi=0)} = \left|
\frac{g_{eff}^{\phi}(\xi)}{g_{eff}^{\phi}(0)} \right|^2 \approx \left( 1 - 4.5\xi\mathcal{A}_{SM} \right)^2.
\end{equation}

Table~\ref{tab:xs_enhancement} illustrates the magnitude of this effect. We observe that for a moderate mixing of $\xi=0.2$, the signal is enhanced by a factor of nearly $47$. For a strong mixing scenario ($\xi=0.5$), the cross-section is enhanced by a factor of over $240$. This quadratic dependence explains why the Radion is potentially observable at the LHC in the constructive interference region, despite the smallness of the gravitational scale $\Lambda_r^{-1}$ relative to the electroweak scale $v^{-1}$.

\begin{table}[h]
\centering
\begin{tabular}{|c|c|c|}
\hline
{Mixing Parameter} & {Coupling Amplitude Ratio} & {Cross-Section Enhancement} \\
{($\xi$)} & {($g_{mix} / g_{pure}$)} & {($\sigma_{mix} / \sigma_{pure}$)} \\
\hline
\hline
$\xi = 0$ (Pure Gravity) & $1$ & $1$ \\
\hline
$\xi = 0.2$ & $\approx 6.9$ & $\approx 47$ \\
\hline
$\xi = 0.5$ & $\approx 15.6$ & $\approx 243$ \\
\hline
\end{tabular}
\caption{Enhancement of the Radion signal strength due to Higgs-mixing interference. The middle column shows the increase in the effective coupling constant, while the right column shows the resulting multiplicative factor on the light-by-light scattering cross-section relative to the $\xi=0$ baseline.}
\label{tab:xs_enhancement}
\end{table}

\subsection{Translation of Experimental Limits}

In order to translate the expected exclusion limits obtained for an ALP ($\Lambda_{ALP}^{lim}$) for a luminosity of $300$ fb$^{-1}$ \cite{Baldenegro:2018hng,Schoeffel:2020svx} into limits for the Radion ($\Lambda_r^{lim}$), we equate their effective couplings. Let us note that these limits are derived under the same assumption that the branching ratio or ALP or Radion in two photons is unity, namely
${\cal B}(a, \phi \rightarrow \gamma \gamma)=1$. 
Adopting this photophilic benchmark facilitates the translation of experimental upper limits on $\sigma \times \mathcal{B}$ directly into the effective coupling plane, treating the ALP and Radion on equal footing regarding their decay topology. This assumption allows us to perform a direct, model-independent comparison of the production cross-sections, isolating the sensitivity to the fundamental scale $\Lambda$ from the complex dependence of the Radion's total width on the hidden sector spectrum.

This yields the conversion formula:
\begin{equation}
    \Lambda_r^{lim} \approx \Lambda_{ALP}^{lim} \times \left( \frac{2\alpha_{EM}}{3\pi} \right) \times \left|
1 - 4.5\xi\mathcal{A}_{SM} \right|.
\end{equation}

Table~\ref{tab:lambda_limits} summarizes the conversion factors for representative values of the non-minimal coupling $\xi$. We verify that for the pure gravity case ($\xi=0$), a sensitivity to a $1$~TeV ALP corresponds merely to a $\sim 1.5$~GeV Radion, which is phenomenologically negligible. However, the introduction of mixing significantly recovers the sensitivity. For $\xi=0.5$, the limit reaches $\sim 24.2$~GeV for the same experimental benchmark.

\begin{table}[h]
\centering
\begin{tabular}{|c|c|c|c|}
\hline
{Mixing} & {Signal Enhancement} & {Scale Factor} & {Projected Radion Limit} \\
{Parameter ($\xi$)} & {(Approx. Cross-section)} & {($\Lambda_r / \Lambda_{ALP}$)} & {(Given $\Lambda_{ALP}^{lim} = 1$ TeV)} \\
\hline
\hline
$\xi = 0$ (Pure) & $1$ (Reference) & $\sim 1/645$ & $\sim 1.5$ GeV \\
\hline
$\xi = 0.2$ & $\times 47$ & $\sim 1/94$ & $\sim 10.6$ GeV \\
\hline
$\xi = 0.5$ & $\times 243$ & $\sim 1/41$ & $\sim 24.2$ GeV \\
\hline
\end{tabular}
\caption{Translation of experimental sensitivity from ALP models to the RS Radion model. The "Scale Factor" indicates how much the limit on $\Lambda$ is reduced compared to an ALP limit due to the loop-suppression factor $2\alpha_{EM}/3\pi$, partially compensated by the mixing enhancement.}
\label{tab:lambda_limits}
\end{table}

\begin{figure}[h]
\centering
\includegraphics[width=0.85\textwidth]{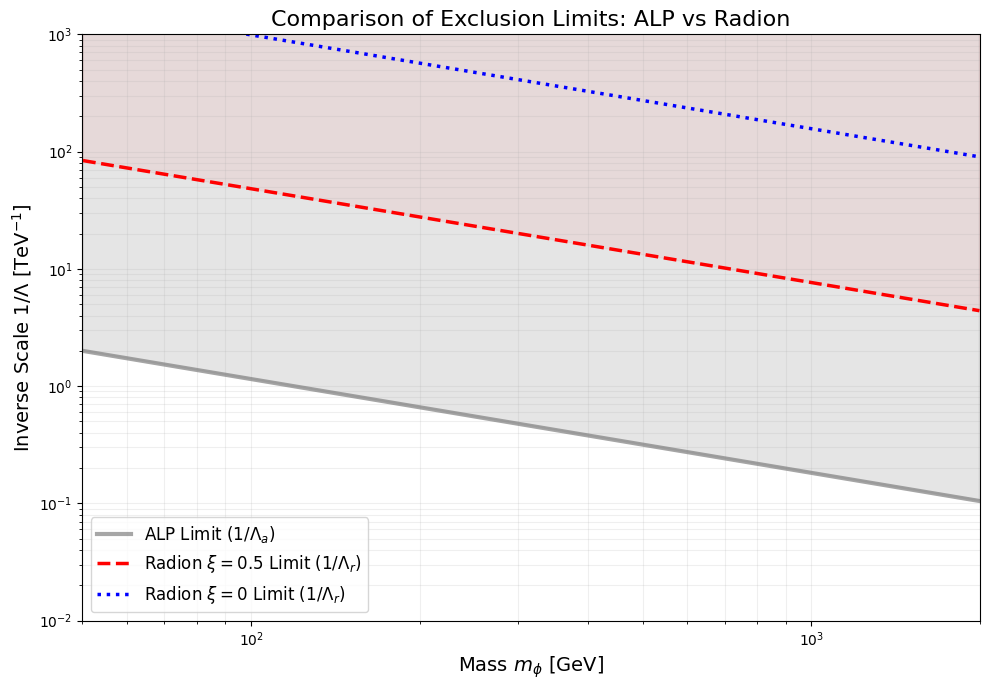}
\caption{Projected exclusion limits on the inverse symmetry breaking scale $1/\Lambda$ as a function of mass,
assuming ${\cal B}(a, \phi \rightarrow \gamma \gamma)=1$. The grey area represents the region excluded for Axion-Like Particles (ALP) by current LHC light-by-light scattering searches. The red dashed line shows the translated limit for a Radion with mixing parameter $\xi=0.5$, and the blue dotted line shows the limit for a pure Radion ($\xi=0$). Higher values on the Y-axis correspond to stronger couplings (smaller $\Lambda$); thus, curves higher up indicate weaker experimental sensitivity to the scale $\Lambda$.}
\label{fig:limit_comparison}
\end{figure}

The resulting constraints are visualized in Figure~\ref{fig:limit_comparison}. The plot displays the inverse scale $1/\Lambda$ as a function of the resonance mass. The grey shaded region indicates the excluded parameter space for an ALP based on current LHC sensitivity. The red dashed line represents the corresponding exclusion limit for a Radion with constructive mixing ($\xi=0.5$), which lies significantly higher in the plot (indicating a weaker limit on $\Lambda$). The blue dotted line for the pure Radion ($\xi=0$) is shifted by nearly three orders of magnitude, illustrating the extreme difficulty of constraining the pure gravitational coupling at the LHC.

\subsection{Note on High-Mass Behavior}
It is crucial to note that the sensitivity recovery shown above relies on the non-decoupling nature of the mixing angle, $\theta \propto m_\phi^2/(m_\phi^2 - m_h^2)$, as derived in Section 3 (see also Ref.~\cite{Csaki:2007ns}). If one were to adopt a decoupling basis where $\theta \propto m_h^2/m_\phi^2$, the enhancement factor would vanish for heavy radions ($m_\phi \gg m_h$), causing the limits to collapse back to the pure gravity case ($\sim 1.5$ GeV) for radion masses in the TeV range.

\section{Conclusion}
\label{sec:conclusion}

In this work, we have revisited the phenomenology of the Randall-Sundrum Radion in the context of light-by-light scattering searches at the LHC. By deriving the effective field theory from the fundamental 5D action, we have highlighted the specific structure of the Radion-photon coupling, which arises solely from the trace anomaly of the energy-momentum tensor. Our analysis explicitly shows that this coupling suffers from a loop-suppression factor of $\alpha_{EM}/6\pi$ compared to the standard dimension-5 operators typical of Axion-Like Particles (ALPs). Consequently, we demonstrated that for a pure gravity dual ($\xi=0$), the Radion signal is suppressed by orders of magnitude relative to an ALP of the same energy scale, making it effectively invisible to current searches. However, we have shown that the phenomenology is drastically altered by the Higgs-Radion kinetic mixing parameter $\xi$. In the regime of constructive interference ($\xi > 0$), the Radion acquires a Higgs-like coupling component that scales with the electroweak scale $v$ rather than the warped scale $\Lambda_r$. This mechanism enhances the production cross-section by factors exceeding $\mathcal{O}(200)$, recovering the sensitivity of the LHC. Using recent high-mass exclusive diphoton searches in proton-proton collisions, we have translated the current exclusion limits on ALPs into the Radion parameter space. We conclude that while a pure Radion remains unconstrained by current data, scenarios with moderate non-minimal coupling ($\xi \gtrsim 0.2$) are beginning to be probed in the hundreds of GeV mass range. These results emphasize the necessity of interpreting light-by-light scattering data not just in terms of effective photon couplings, but within the complete mixing landscape of the scalar sector.

 
 {\bf Data availability} \\
No data was used for the research described in the article.

{\bf Declaration of interests} \\
The authors declare that they have no known competing financial interests or personal relationships that could have appeared to influence the work reported in this paper.

\newpage

\section{Appendix}

Let us prove that a fundamental mass  scale  $m_0 \sim M_{Pl}$  on the Planck-brane ($y=0$) is perceived on the TeV-brane ($y=y_c$) as an exponentially smaller warped mass
$$
m_{\text{phys}} = m_0 \times e^{-ky_c},
$$
which is the key formula to move the vacuum expectation  value or the Higgs mass to the electroweak scale on the Tev-brane. Let us consider the Lagrangian density for the Higgs field on the TeV-brane
$$
\mathcal{L}_{\text{brane}} = \sqrt{-g_{\text{brane}}} \left[ g^{\mu\nu}_{\text{brane}} (\partial_\mu H)^\dagger (\partial_\nu H) - V(H) \right],
$$
where $g^{\mu\nu}_{\text{brane}}$ is the RS metric induced on the TeV-brane and $V(H)$ is the Higgs potential
 $$
V(H) = -m_0^2 |H|^2 + \lambda |H|^4 + m_0^4/(4\lambda)+...
$$
Here $m_0$ is the mass parameter, with $m_0 \sim M_{Pl}$ at $y=0$. Using the RS metric
$ds^2 = e^{-2ky} \eta_{\mu\nu} dx^\mu dx^\nu + dy^2$, we can compute
the induced metric components $g_{\mu\nu}^{\text{brane}}$, namely
$$
g_{\mu\nu}^{\text{brane}}(x) = e^{-2ky_c} \eta_{\mu\nu}.
$$
Its inverse is thus 
$$
g^{\mu\nu}_{\text{brane}}(x) = e^{2ky_c} \eta^{\mu\nu}.
$$
We also have
$$\sqrt{-g_{\text{brane}}} = \sqrt{-\det(e^{-2ky_c} \eta_{\mu\nu})} = \sqrt{-(e^{-2ky_c})^4 \det(\eta_{\mu\nu})} = e^{-4ky_c}$$
(since $\sqrt{-\det(eta_{\mu\nu})} = 1$ for the flat Minkowski metric $\eta_{\mu\nu}$). Them, we can write
$$
\mathcal{L}_{\text{brane}} = e^{-4ky_c} \left[ (e^{2ky_c} \eta^{\mu\nu}) (\partial_\mu H)^\dagger (\partial_\nu H) - (-m_0^2 |H|^2 + \lambda |H|^4 + m_0^4/(4\lambda)) \right],
$$
or
$$
\mathcal{L}_{\text{brane}} = e^{-2ky_c} \eta^{\mu\nu} (\partial_\mu H)^\dagger (\partial_\nu H) + e^{-4ky_c} m_0^2 |H|^2 - e^{-4ky_c} \lambda |H|^4-e^{-4ky_c} m_0^4/(4\lambda).
$$
This Lagrangian density doesn't look like a standard 4D Lagrangian density  because  the kinetic term has that $e^{-2ky_c}$ factor. In order to compare it with the SM and identify the physical mass, we need to rescale the Higgs field $H$. Let us define a rescaled field $\tilde{H}$ such that
$$
H = e^{ky_c} \tilde{H}.
$$
Then
$$
e^{-2ky_c} \eta^{\mu\nu} (\partial_\mu (e^{ky_c}\tilde{H}))^\dagger (\partial_\nu (e^{ky_c}\tilde{H})) = e^{-2ky_c} e^{2ky_c} \eta^{\mu\nu} (\partial_\mu \tilde{H})^\dagger (\partial_\nu \tilde{H}) = \eta^{\mu\nu} (\partial_\mu \tilde{H})^\dagger (\partial_\nu \tilde{H}).
$$
This corresponds to the standard kinetic term, expressed with the rescaled field $\tilde{H}$. Then, this is clear that the mass term reads as
$$
 e^{-4ky_c} m_0^2 |H|^2= (m_0 e^{-ky_c})^2 |\tilde{H}|^2.
$$
Finally, we get
$$
\mathcal{L}_{\text{brane}} = \eta^{\mu\nu} ( \partial_\mu \tilde{H})^\dagger (\partial_\nu \tilde{H}) +(m_0 e^{-ky_c})^2 |\tilde{H}|^2 - \lambda |\tilde{H}|^4+cte.
$$
Notice the quartic coupling $\lambda$ is unchanged by the rescaling.
This implies that
$$m_{\text{phys}}^2 = (m_0 e^{-ky_c})^2,$$
which concludes the proof.
\newpage

\providecommand{\href}[2]{#2}\begingroup\raggedright

\endgroup

\end{document}